%% file: usenix2025_SOUPS.tex
\documentclass[letterpaper,twocolumn,10pt]{article}
\usepackage{usenix2025_SOUPS}

\usepackage{filecontents}
\usepackage{graphicx}

\AtBeginEnvironment{quote}{\itshape}

\begin{document}

\date{}

\title{\Large \bf Intergenerational Support for Deepfake Scams Targeting Older Adults}

\def\plainauthor{}

\author{
{\rm Karina LaRubbio}\\
Brown University
\and
{\rm Alyssa Lanter}\\
Brown University
\and
{\rm Seihyun Lee}\\
Tenafly High School
\and
{\rm Mahima Ramesh}\\
Acton-Boxborough Regional High School
\and
{\rm Diana Freed}\\
Brown University
} 

\maketitle
\thecopyright

\begin{abstract}
AI-enhanced scams now employ deepfake technology to produce convincing audio and visual impersonations of trusted family members, often grandchildren, in real time. These attacks fabricate urgent scenarios, such as legal or medical emergencies, to socially engineer older adults into transferring money. The realism of these AI-generated impersonations undermines traditional cues used to detect fraud, making them a powerful tool for financial exploitation. In this study, we explore older adults’ perceptions of these emerging threats and their responses, with a particular focus on the role of youth, who may also be impacted by having their identities exploited, in supporting older family members’ online safety. We conducted focus groups with 37 older adults (ages 65+) to examine their understanding of deepfake impersonation scams and the value of intergenerational technology support. Findings suggest that older adults frequently rely on trusted relationships to detect scams and develop protective practices. Based on this, we identify opportunities to engage youth as active partners in enhancing resilience across generations.
\end{abstract}

\input{sections/0_intro}

\input{sections/2_method}
\input{sections/3_result}

\input{sections/4_discussion}

\bibliographystyle{plain}
\bibliography{references}

\end{document}

%% file: sections/0_intro.tex
\section{Introduction}
In the United States, older adults lost an estimated \$61.5 billion to fraud in 2023 \cite{lina_m_khan_protecting_2024}. Advances in technology have given rise to more sophisticated scams. In the "grandparent scam," attackers impersonate trusted family members in need of urgent financial support, sometimes using AI-generated deepfakes, such as audio or visual content \cite{zhai_hear_2025}. These scams have recently prompted media coverage\cite{vellani_her_2025, alfonsi_how_2023, tresiman_dozens_2025} and multiple consumer alerts \cite{noauthor_fcc_2025,noauthor_fincen_2024}, emphasizing their rising prevalence. 

Deepfake impersonation scams deviate from the threat models for other scams because they create a situation of \textbf{dual victimization}\footnote{We recognize there are different terms used to refer to people experiencing the negative impacts of online safety threats, such as "survivor". We use the term "victim" to align with the criminal nature of these offenses and participant preferences.}:
in a successful attack, the older adult being targeted experiences a monetary loss, and the family member being impersonated unknowingly has their likeness used to cause harm.  Older adults are not inherently more vulnerable to scams than other demographics \cite{morrison_recognising_2023}, though they are specifically targeted in this context. Intergenerational support strategies are especially relevant in this context, where two parties are victimized, older adults and youth. Prior work emphasizes the importance of community support such as peers in mitigating scams against older adults \cite{nicholson_if_2019, leedahl_implementing_2019}, sometimes identifying younger family members as crucial sources of support \cite{deng_auntie_2025, tang_i_2022,murthy_individually_2021}. 

This research aims to understand older adults' perceptions of deepfake impersonation scams, specifically focusing on the role of youth supporters, such as grandchildren. We address the research questions: 1) How do older adults currently perceive and protect themselves against deepfake impersonation scams? 2) What opportunities exist to engage youth to support older adults against deepfake impersonation scams?

Through focus groups with older adults, we offer insights into how older adults perceive and respond to deepfake impersonation scams, contributing a greater understanding of their reliance on interpersonal networks for scam detection and response. We additionally contribute a dual victimization framework for deepfake impersonation scams, centering youth as both affected parties and potential allies. Finally, we identify actionable opportunities to foster intergenerational digital resilience, including youth-driven support strategies. Motivated by participant perspectives, these approaches seek to address the evolving threat of AI-enabled scams.

%% file: sections/2_method.tex
\section{Methods}
We conducted six focus groups with older adults (n = 37) aged 70 to 94 years in an independent living facility in the northeast United States between February and March 2025. The focus groups lasted approximately 60 minutes and ranged in size from six to ten participants. Participants received \$10 as compensation. Discussion guides included questions on perceptions of AI technology, online safety, and deepfake impersonation scams specifically. Each focus group was audio recorded with participants' consent and transcribed for qualitative analysis. This study was approved by our IRB. 

%% file: sections/3_result.tex
\section{Results}

After discussing what deepfake impersonation scams are and how they work, our participants expressed broad familiarity with these threats. However, some were skeptical about the fidelity of deepfake voices, expressing that they ``couldn't believe [they] would fall for such a thing'' (P22). Others, and especially those who had experienced financial scams in the past, expressed fear about being scammed again. About distinguishing real callers from attackers, one participant stated:
\begin{quote}
    ``You really can't tell... I just think I'm going to hit the wrong button at some time, and it's not going to be good. So I'm really nervous about that.'' (P12)
\end{quote}
Many expressed that their concerns were heightened by AI technology's ability to generate synthetic content, causing them to ``second guess things'' (P3). In response to the threats our participants perceived from possible deepfake impersonation scams, many described collaborating with younger family members to access support and build intergenerational protective practices in the following ways:

\textbf{Avoidance.} For many of our participants, their younger family members advised them to avoid real-time interactions with unknown parties altogether. One participant shared:
\begin{quote}
    ``My grandson's advice that I've used constantly is: if it's that important, they'll leave a message.'' (P7)
\end{quote}
While this strategy did support their resilience against scams, our participants expressed that ignoring unknown callers completely led them to miss important information, such as time-sensitive medical notifications. 

\textbf{Contact for verification.} Many participants expressed that they already relied on making direct contact with trusted family members to ensure that information they received from a suspicious source was reliable, and would continue this practice should they encounter a deepfake impersonation scam. One participant suggested:
\begin{quote}
    ``I'd ask the caller, who else in our family knows about this and then ... call them myself.'' (P32)
\end{quote}
However, some expressed that this would be ineffective if they could not immediately contact their family members. 

\textbf{Shared knowledge.} Many participants expressed that they had already collaborated with family members to create code words that only family members knew, which could be used in situations of uncertainty. Others suggested asking a question that only family members would know the answer to in order to verify the caller's identity. One participant recounted their strategy for avoiding a recent impersonation scam attempt they had experienced:
\begin{quote}
    ``I said, what's your name? He said, grandma, you don't know my name? I said, well I have three grandsons, so which one are you? And then he hung up. Just ask a family question that they won't know.'' (P27)
\end{quote}
Participants thought that using shared knowledge was a strong protective practice. However, this requires them to recognize that a call is suspicious, which may become increasingly challenging with advances in deepfake technology.

%% file: sections/4_discussion.tex
\section{Discussion}
The protective practices our participants suggested for deepfake impersonation scams all relied on support from younger family members to some extent. Advice from family members to avoid contact with attackers, reach out to verify information, and refer to shared knowledge was helpful because it was provided proactively, before the older adult was targeted. Returning to the dual victimization framework, youth may have an interest in protecting their likeness from use in deepfake impersonation scams that cause harm to their older family members. Motivated by participants' insights and perspectives from our youth team members, we considered opportunities to encourage youth engagement in their older adult family members' online safety:

\textbf{In-Class Online Safety Education.}
Interventions targeting youth's own online safety have been integrated into educational curriculums with success \cite{freed_understanding_2023}. Building on this, curriculums could explain how youth can support their older adult family members' online safety. Specifically, programming could introduce youth to collaborative protective practices, such as agreeing on a secret code word, prompting them to proactively engage in these conversations.

\textbf{Leveraging Social Media.} Youth are often exposed to information through social media \cite{alluhidan_teen_2024}. In the context of deepfake impersonation scams, using social media to increase awareness and share guidelines on how to talk to older adult family members about scams could engage broad youth audiences.

Future work should investigate how youth perceive their role in older adult family members' online safety and respond to the suggested interventions. As previous work has surfaced, intergenerational technology support can create challenging family dynamics \cite{murthy_individually_2021}, which future work should also explore from youth perspectives. As scams become more technically sophisticated and involve older adults' family members, innovative approaches are needed to engage these stakeholders and build resilience across generations.